\documentclass{aa}

\usepackage[]{graphics}
\usepackage[]{rotating}
\usepackage[]{psfig}

\begin{document}

\thesaurus{03(11.04.4;11.06.2;11.16.1;04.03.1)}

\title{The central region of the Fornax cluster -- I. A catalog and photometric 
properties of galaxies in selected CCD fields}

\author {Michael Hilker \inst{1,4} \and Markus Kissler-Patig \inst{2,1,3} \and 
Tom Richtler \inst{1} \and Leopoldo Infante \inst{4} \and Hernan Quintana 
\inst{4}
} 

\offprints {M.~Hilker}
\mail{mhilker@astro.uni-bonn.de}

\institute{
Sternwarte der Universit\"at Bonn, Auf dem H\"ugel 71, 53121 Bonn, Germany
\and
UCO/Lick Observatory, University of California, Santa Cruz, CA 95064, USA
\and
Feodor Lynen Fellow of the Alexander von Humboldt Foundation
\and
Departamento de Astronom\'\i a y Astrof\'\i sica, P.~Universidad Cat\'olica,
Casilla 104, Santiago 22, Chile
}

\date {Received --- / Accepted ---}

\titlerunning{The central region of the Fornax cluster -- photometry of galaxies}
\authorrunning{M.~Hilker et al.}
\maketitle

\begin{abstract}

We present a photometric catalog (based on $V$ and $I$ photometry)
of galaxies in the central regions of the Fornax galaxy cluster.
Our 11 CCD fields cover 0.17 square degrees in total.
The limiting surface brightness is around 24 mag arsec$^{-2}$, similar 
to that of Ferguson's (\cite{ferg}) catalog,
whereas our limiting total magnitude is around $V\simeq22$ mag, about
two magnitudes fainter. It is the surface brightness limit, however,
that prevents us from detecting the counterparts of the faintest Local
Group dwarf spheroidals. The photometric properties of all objects are
presented as a catalog (Appendix A)\footnote{The tables of Appendix A and 
Appendix B are only available in electronic form at the CDS via anonymous
ftp to cdsar.u-strasbg.fr (130.79.128.5) or via 
http://cdsweb.u-strasbg.fr/Abstract.html}. 
The properties and fit parameters of the
surface brightness profiles for a sub-sample are presented as a second
catalog (Appendix B)$^1$.

We can only add 4 new dwarf galaxies to Ferguson's catalog.
However, we confirm that the dwarf galaxies in Fornax follow a similar surface
brightness -- magnitude relation as the Local Group dwarfs. They
also follow the color (metallicity) -- magnitude relation seen in other galaxy
clusters. A formerly suspected excess of dwarf galaxies surrounding the
central giant cD galaxy NGC~1399 can finally be ruled out.

An enhanced density of objects around NGC 1399 can indeed be seen, but it
appears displaced with respect to the central galaxy and is identified
as a background cluster at $z=0.11$ in Paper II of these series, which
will discuss spectroscopic results for our sample. 
                                               

\end{abstract}


\section{Introduction}

The galaxy population in the Fornax cluster is well studied down
to a B magnitude of 19 mag by the photographic survey of
Ferguson (\cite{ferg}) which resulted in the Fornax Cluster Catalog (FCC). i
In addition, Davies et al.~(\cite{davi}) and Irwin et al.~(\cite{irwi})
provided a catalog of the (very)
low surface brightness galaxies in this region, which then also have been 
studied by follow-up
CCD photometry (Bothun et al.~\cite{both}, Cellone et al.~\cite{cell96}).

The photographic data of Ferguson (\cite{ferg}) were obtained on the same 
telescope
as the present study. In Ferguson's survey a diameter limit of $17\arcsec$ was
chosen to reduce the number of background galaxies. In our CCD survey
galaxies with smaller scale lengths were included in order to search for
compact dwarf galaxy, e.g. similar to M32, and to extend the galaxy counts
to fainter absolute magnitudes. Furthermore, CCD photometry provides the 
possibility to search for objects in the
immediate environment of giant galaxies (where the galaxy light dominates) when
adopting appropriate reduction techniques.

\begin{figure*}
\begin{rotate}{-90}
\end{rotate}
\vspace{11.2cm}
\caption{Image taken from the Digital Sky Survey, showing the observed CCD
fields in the Fornax cluster. Field B2 lies outside this image to the east}
\end{figure*}

Our main goal is to find dwarf galaxies in the
vicinity of giant ellipticals in the Fornax cluster, especially the central 
galaxy NGC~1399. One striking characteristic of NGC~1399 (and other central 
galaxies
in clusters) is its extraordinarily rich globular cluster system (GCS), whose 
origin
is a matter of lively debate. Several authors investigated photometric and
spectroscopic properties of the GCS (e.g. Hanes \& Harris \cite{hane}; Bridges 
et al.~\cite{brid}; Wagner et al.~\cite{wagn}; Grillmair et al.~\cite{gril};
Kissler-Patig et al.~\cite{kiss97}, \cite{kiss98a},b; Forbes et 
al.~\cite{forb}).
Further, NGC~1399 possesses an extended cD halo 
(Schombert \cite{scho}; Killeen \& Bicknell \cite{kill}), which
seems to be typical for the brightest cluster galaxies of dynamically evolved
clusters (L\'opez-Cruz et al.~\cite{lope}). 
The building-up of such a rich GCS as well as the formation
of a cD halo are not well understood. Both properties might be related to the 
infall of dwarf galaxies in the cluster center. A more detailed discussion on 
this topic will be given in Paper III of this series (Hilker et 
al.~\cite{hilk98iii}).

In the present paper the photometric properties of all galaxies
found in the observed fields are cataloged. The identification, classification
and the photometry of the objects as well as the completeness of our counts
are described.
Furthermore, we present the spatial and the color distribution of the galaxies 
as well as the 
surface brightness profile parameters of a subsample of galaxies with a
sufficient resolution in their angular diameters. In this paper we adopt
a distance to the Fornax cluster of 18.2 Mpc or $(m-M)_0 = 31.3$ mag
(Kohle et al.~\cite{kohl}, recalibrated with new distances of Galactic GCs,
Gratton et al.~\cite{grat}, following Della Valle et al.~\cite{dell}).
The spectroscopy and the determination of radial velocities of the brightest
galaxies in our sample represents the topic of a second paper (Hilker et 
al.~\cite{hilk98ii}, hereafter Paper~II).
%
%
\section{Selection of the observed fields}

\subsection{Observed fields}

We selected 11 fields in the Fornax cluster which comprise giant galaxies 
as well as control fields without any bright Fornax galaxy.
A mosaic of 4 fields matches the central giant 
elliptical NGC~1399 (named F1, F2, F3, and F4 as shown in Fig.~1).
Five other target galaxies are the giant ellipticals NGC~1374, NGC~1379,
and NGC~1427, the S0 galaxy NGC~1387, and the irregular NGC~1427A. The 2 
additional fields (named B1 and B2 in the following) are located 
13$\arcmin$ and 95$\arcmin$ east of NGC~1399, respectively.
Note that field B2 is not suitable for an estimation of the absolute 
galaxy background, since it still lies in a region
where Fornax member galaxies contribute to the galaxy counts, as seen in the 
radial density profiles of Ferguson (\cite{ferg}). All fields, except B1 and 
the four NGC~1399 fields, can serve as relative
backgrounds to identify a possible excess population near NGC~1399 with respect
to the ``normal'' Fornax field galaxies.
In total, our fields cover an area of about 600 square arcmin (or 0.17 square
degree).
Figure 1 is an extraction of the Digital Sky Survey and shows the location
of all our fields except B2.

In a second observing run, 2 additional outlying background fields $10\degr$
north (B3) and $15\degr$ south of NGC~1399 have been observed (see Sect.~3.1).

\subsection{Nomenclature of catalog fields}

In the following the acronym CGF (Catalog of Galaxies in Fornax) will be used 
for the nomenclature of galaxies in the observed
fields followed by a sequence
number of the field and the sequence number of the galaxy in this field
ordered by decreasing magnitude. For example, CGF~5-12 is the 12th brightest
galaxy in field 5. The sequence numbers of the fields are ordered with
increasing distance to the center of the Fornax cluster. The
cross references to the CCD fields as described above are (see also
Fig.~1): CGF~1 = F1 (NGC 1399, SE field), CGF~2 = F2 (NGC 1399,
NE field), CGF~3 = F3 (NGC 1399, SW field), CGF~4 = F4 (NGC 1399, NW field),
CGF~5 = B1, CGF~6 = NGC~1387, CGF~7 = NGC~1427A, CGF~8 = NGC~1379,
CGF~9 = NGC~1374, CGF~10 = NGC~1427, and CGF~11 = B2.

\section{Observations and Reduction}

The observations were performed with the 2.5m DuPont telescope at Las 
Campanas Observatory, Chile, in the nights of 26-29 September, 1994. 
A Tektronix $2048\times 2048$ pixel chip has been used, with a pixel size of 
$21\mu$m or 
$0\farcs227$ at the sky, corresponding to a total field of view of 
$7\farcm7\times7\farcm7$.

Deep exposures in Johnson V and Cousins I were acquired. Typical total 
exposure
times are between 30 min and 60 min per field in both filters. The seeing,
given as the full width at half maximum (FWHM) of a Gaussian profile, 
ranged between $0\farcs9$ and $1\farcs5$.
The exposures of the south-west field of the NGC~1399 mosaic were affected
by a tracking jump in one long V exposure and the corruption of one long
I exposure. Omitting these two exposures the total exposure time was reduced to 
15 min in both filters.
Nevertheless, this field has been considered in several aspects of
our investigations since it contains some interesting bright galaxies. 
For the photometric calibration typically 15--30 standard stars from the 
Landolt (\cite{land}) list have been observed throughout the nights.

Table 1 summarizes the observations. In addition, the detection threshold 
$\mu_{limit}$ in mag~arcsec$^{-2}$ in $V$ and $I$ is given for each field
(see also Sect.~4).
For more information about the calibration see Kissler-Patig 
et al.~(\cite{kiss97}).

\begin{table}
\caption{The observations of the Fornax fields obtained with the 2.5m telescope
and the background fields obtained with the 1.0m telescope on Las Campanas.
In addition, the surface brightness threshold $\mu_{limit}$ in 
mag~arcsec$^{-2}$, above which objects of at least five connected pixels are 
detected, is given for each field}
\begin{flushleft}
\begin{tabular}{llll@{\hspace{3mm}}c@{\hspace{3mm}}c}
\hline\noalign{\smallskip}
\multicolumn{2}{l}{Field \& Filter} & Date & Exp. time & Seeing & $\mu_{limit}$\\
 & & & [s] & [$\arcsec$] & [mag/$\sq \arcsec$] \\
\noalign{\smallskip}
\hline\noalign{\smallskip}
{\small NGC~1374} & $V$  & 28.9.94 & $2 \times 1200$ & 1.2 & 25.0\\
         & $I$ & 29.9.94 & $1200$  & 1.5 & 24.2\\
	 &     &         & $+ 2\times 600$ & & \\
{\small NGC~1379} & $V$  & 29.9.94 & $2\times 1200$ & 1.2 & 25.4\\
         & $I$ & 26.9.94 & $3\times 1200$ & 1.4 & 24.7\\
{\small NGC~1387} & $V$  & 28.9.94 & $2\times 900$ & 1.3 & 24.8\\
         & $I$ & 28.9.94 & $2\times 900$ & 1.2 & 23.5\\
{\small NGC~1427} & $V$  & 26.9.94 & $3\times 1200$ & 1.5 & 25.3\\
         & $I$ & 26.9.94 & $3\times 1200$ & 1.3 & 24.6\\
{\small NGC~1427A} & $V$  & 28.9.94 & $2\times 900$ & 0.9 & 25.1\\
         & $I$ & 28.9.94 & $2\times 900$ & 0.9 & 24.0\\
{\small NGC~1399 F1} & $V$  & 27.9.94 & $2\times 900$ & 0.9 & 24.7\\
         & $I$ & 27.9.94 & $2\times 900$ & 1.1 & 24.1\\
{\small NGC~1399 F2} & $V$  & 27.9.94 & $2\times 900$ & 0.9 & 24.8\\
         & $I$ & 27.9.94 & $2\times 900$ & 1.3 & 24.1\\
{\small NGC~1399 F3} & $V$  & 28.9.94 & $1\times 900$ & 1.2 & 24.3\\
         & $I$ & 28.9.94 & $1\times 900$ & 1.2 & 22.1\\
{\small NGC~1399 F4} & $V$  & 28.9.94 & $2\times 900$ & 1.0 & 24.9\\
         & $I$ & 28.9.94 & $2\times 900$ & 1.1 & 24.3\\
{\small B1} & $V$  & 29.9.94 & $2\times 900$ & 1.1 & 25.2\\
         & $I$ & 29.9.94 & $2\times 900$ & 1.1 & 24.0\\
{\small B2} & $V$  & 29.9.94 & $2\times 900$ & 1.0 & 25.3\\
         & $I$ & 29.9.94 & $2\times 900$ & 1.2 & 24.0\\
 & & & & & \\
{\small B3} & $V$  & 2.12.96 & $3\times 1200$ & 1.5 & 26.0\\
      & $I$ & 2.12.96 & $3\times 1200$ & 1.5 & 24.5\\
{\small B4} & $V$  & 2.12.96 & $3\times 1200$ & 1.6 & 25.5\\
    & $I$ & 2.12.96 & $3\times 1200$ & 1.7 & 24.2\\
\noalign{\smallskip}
\hline
\end{tabular}
\end{flushleft}
\end{table}

Bias subtraction, flat-fielding, long exposures combination, modelling, and
subtraction of the galaxy light of the giants were done with IRAF (for details
see  Kohle et al.~\cite{kohl} and Kissler-Patig et al.~\cite{kiss97}). After 
these procedures
frames with flat backgrounds ready for object search and photometry were
obtained.

\subsection{Background fields}

To get an estimation of the absolute background galaxy counts 
two additional fields $10\degr$  north (B3) and $15\degr$ south (B4) of 
NGC~1399 have been observed. The total magnitudes and colors of the galaxies
have been determined for a statistical comparison with the photometric 
properties of galaxies
in the Fornax fields. The surface brightness (SB) profiles have not been
measured, because the 
resolution of the images is 3 times lower than in the run at the 2.5m telescope.

The observations have been performed in a second run with the 1.0m Swope
telescope at Las
Campanas Observatory, Chile, in the night of 2/3 December, 1996. 
A SITe\#1 $2048\times 2048$ pixel chip has been used, with a pixel size
$0\farcs694$ at the sky, corresponding to a total field of view of
$23\farcm69\times23\farcm69$. Table 1 (at the bottom) gives an observation log 
of the V and
I exposures. As in the first run the photometric calibration was done via
standard stars from the Landolt (\cite{land}) list.

\section{Object detection and photometry}

The identification, photometry and classification of the objects have been
performed with the program SExtractor (Source-Extractor, Bertin 
\cite{bert95}).

First, a background map was constructed and subtracted from the image.
The sky map is a bilinear interpolation between the meshes of a grid with
a mesh size of $56 \times 56$ pixel (or $128 \times 128$ pixel in the case of
the more extended Fornax dwarf ellipticals). In addition, a median filter
of $5 \times 5$ pixel has been applied in order to suppress possible
overestimations due to bright stars.
Then the image was convolved with a Gaussian with a FWHM slightly larger
than that for stellar images in order to favour the detection of marginally
resolved objects. A FWHM of $2\arcsec$ was chosen for all fields
except the background fields B3 and B4, which were convolved with a Gaussian
of $1\farcs5$ FWHM.
Objects were found with a threshold of about 2 sigmas above the sky level.
The level of the lowest isophote in $V$ and $I$ above which objects were
detected
is given in Table 1 for the different fields. The minimum number
of connected pixels for a detection was chosen to be 5 in all fields.
Composite objects were deblended by a multithresholding algorithm.
In all fields the identifications have been controlled by eye, obvious
multi-identifications of the same object were removed and objects that had been 
missed
by the finding algorithm were added. The number of added objects
was always below 2\% of the total number.

The finding completeness starts to drop in $V$ at magnitudes between 22.0 and
22.5 mag (except the NGC~1399 SW field: 21.0) and in $I$ at magnitudes between
21.0 and 21.5 mag (NGC~1399 SW field: 20.0) depending on the seeing and 
exposure times of the different fields.
The finding limit for low surface brightnesses varies between 23.0 and 24.0 
mag~arcsec$^{-2}$ peak surface brightness in $V$ ($\mu_{\rm peak}$ is
the surface brightness of the central pixel as given by SExtractor).
This latter limit is about the same as in the FCC, whereas the limiting 
total magnitude is about 2 magnitudes fainter than in the FCC.

The photometry was done via elliptical apertures whose ellipticity and position
angle are defined by the second order moments of the light distribution.
Total magnitudes are computed in two different ways. For isolated objects 
the flux is measured within
an aperture calculated by a further development of Kron's ``first
moment'' algorithm (Kron, \cite{kron}; see also Infante \cite{infa87}). 
For overlapping objects, i.e which have neighbours within the elliptical 
aperture,
the fraction of flux lost by the isophotal magnitude 
(using the detection threshold as the lowest isophote) is estimated and 
corrected assuming that the intensity profiles have gaussian wings
because of atmospheric blurring. 

\begin{figure}
\hspace{0.0cm}\psfig{figure=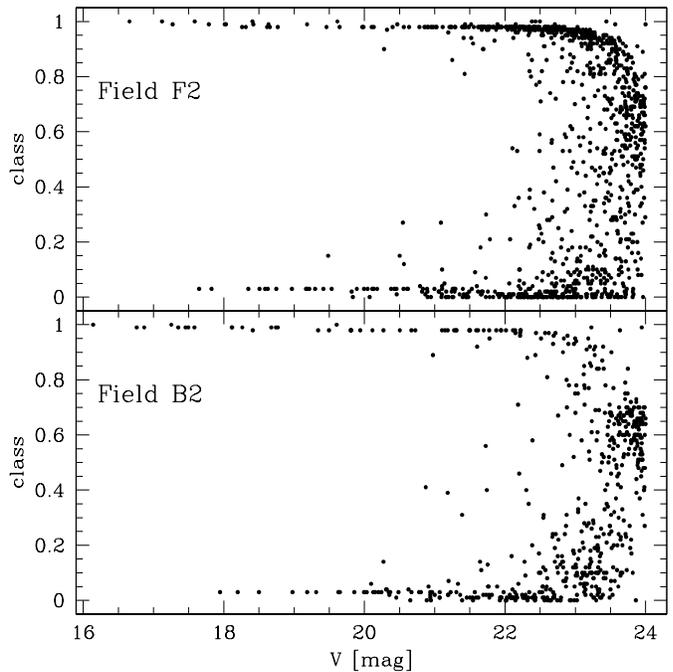,height=8.6cm,width=8.6cm
,bbllx=9mm,bblly=65mm,bburx=195mm,bbury=246mm}
\vspace{0.4cm}
\caption{The classifier value ($0 =$ galaxy, $1 =$ point source) is plotted
versus
the $V$ magnitude for two different CCD fields (see Fig.~1). For $V<21$
the seperation of point sources and galaxies is obvious. For fainter magnitudes
all objects with classifier values below 0.35 ihave been identified as well
resolved.
Note that most point sources in field F1 are globular clusters that belong to
the central galaxy NGC~1399}
\end{figure}

The color determination was done by measuring aperture magnitudes from
circular apertures with $3\arcsec$ diameter in both filters.
Peak surface brightnesses ($\mu_{\rm peak}$ in mag~arcsec$^{-2}$) are 
calculated from the peak intensity at the central pixel.
Note that for objects with apparent core radii intrinsically smaller than the
seeing the derived central surface brightness is a lower limit compared to
the true central surface brightness due to the convolution with the seeing.
A more detailed analysis of the surface brightness 
profiles of the brighter galaxies in our sample is given in Sect.~6.

The photometric calibration was done by applying
the calibration equations for aperture photometry given in
Kissler-Patig et al.~(\cite{kiss97}).

\section{Classification and selection}

The separation of resolved ($=$ galaxies) and unresolved ($=$ stars, globular
clusters, and
unresolved background galaxies) was done with the ``star/galaxy classifier''
developed by Bertin \& Arnouts (\cite{bert96}). It is a neural network 
trained program
that classifies each object with a ``stellarity index'' between 0 (galaxy) and
1 (point source).
In all fields the separation works fine down to a $V$ magnitude of 21 mag.
Beyond this magnitude the classifier values start to scatter. Figure 2 shows
as an example the classifier versus $V$ magnitude plots for the NE field (F2) of
NGC~1399 and the background field B2. Eye control in all fields showed 
that all objects with classifier 
values below 0.35 are well resolved objects, whereas objects above this 
value can not be clearly classified.

Therefore all objects were selected that have classifier
values below 0.35. 
Down to a $V_{\rm tot} = 22.0$ and 23.0 mag we found 873 and 1775 galaxies
respectively.
The properties of all galaxies brighter than 22.0 mag are compiled in a 
catalog, see Appendix A. This cutoff in the final sample was chosen 
for several reasons: the finding completeness starts to drop, the scatter in
the classifier values increases significantly. Also, at this magnitude we
would not find any dEs that follow the
surface brightness -- magnitude
relation of Local Group or Virgo dwarf spheroidals
(e.g. Kormendy \cite{korm}, Binggeli \cite{bing}) due to the limit in surface 
brightness. See Sect.~7 for a discussion of the
completeness of dwarf galaxies in the Fornax distance.

\begin{figure}
\vspace{0.4cm}
\hspace{-0.1cm}\psfig{figure=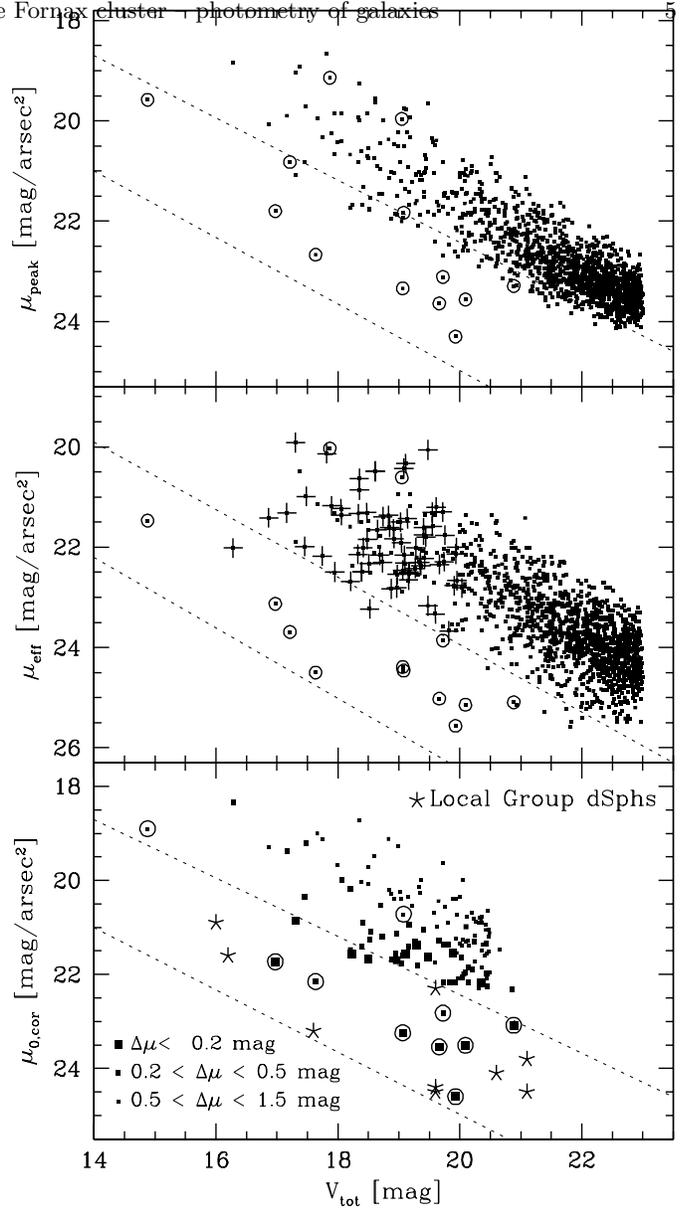,height=14.5cm,width=8.6cm
,bbllx=15mm,bblly=67mm,bburx=115mm,bbury=231mm}
\vspace{0.4cm}
\caption{The three panels show surface brightness versus total 
magnitude plots for all objects in the photometric catalog (small dots).
In the upper two panels $\mu_{\rm peak}$ (not seeing corrected) and
$\mu_{\rm eff}$ from growth curve analyses are plotted. 
In the lower panel the corrected central surface brightness (as described in 
the text) is shown for a subsample of galaxies. $\Delta\mu$ is the
difference between the apparent and corrected central surface brightness.
In all diagrams the encircled dots indicate Fornax cluster members, as follows 
from their radial velocity (see Paper~II) and/or SB profile (Sect.~6).
All other objects that are not encircled are suspected to be background
galaxies. In the middle panel definite background galaxies, according to their 
radial velocities, are symbolised with crosses. In all panels
the dotted lines mark the region that is defined by the $\mu$-$V$ relation
of the Local Group and Virgo dEs (the asterisks in the lower panel are
Local Group dSphs shifted to the Fornax distance; Kormendy \cite{korm}, 
Mateo et al.~\cite{mate}).
}
\end{figure}

In the background fields B3 and B4 we found with the same selection criteria
668 and 1022 down to $V = 22.0$ mag respectively.
However, one has to be careful when comparing these results with those
of the other CCD fields.
The pixel size is about 3 times larger than in the first run leading to
a 9 times higher area covered by each pixel. We simulated this resolution for
two fields of our first run by binning $3 \times 3$ pixel and run SExtractor
again. On the one hand, some galaxies have been classified as point sources due
to their small angular sizes below the ``new'' resolution. On the other hand,
some new ``galaxies'' have been gained due to the overlap of objects very close
in the high resolution image. Down to our magnitude limit of $V = 22.0$
mag loss and gain of galaxies are nearly balanced and in the order of 8\% of
the total galaxy counts.

\section{Surface brightness profiles}

In combination with the morphological appearance of the observed galaxies,
the analysis of their SB profiles provides a reliable tool
to classify them as cluster dwarf
galaxies or background galaxies (e.g. Sandage \& Binggeli \cite{sand}).
In the surface brightness versus magnitude diagram ($\mu$-$V$ diagram)
the dEs follow a well defined sequence.

\begin{table*}
\caption{Photometric properties and profile fit parameters of dwarf galaxies
in Fornax. The classification as dwarf galaxy is based on the morphological
appearance, surface brightnesses, and in four cases on radial velocities.}
\begin{flushleft}
\small{
\hspace*{-0.4cm}\begin{tabular}{lrlcccccrcrcc}
\noalign{\smallskip}
\hline
\noalign{\smallskip}
Id & FCC & Type & RA(2000) & Dec(2000) & $V$ & $(V-I)$ & $\mu_{\rm peak}$ &
$D_{26}$ & $\mu_{\rm eff}$ & $a_{\rm eff}$ & $\mu_{\rm 0,exp}$ & $\alpha$\\
 & & & [$^h$:$^m$:$^s$] & [$\degr:\arcmin:\arcsec$] & [mag] & [mag] & 
[mag/$\sq \arcsec$] & [$\arcsec$] & [mag/$\sq \arcsec$] & [$\arcsec$] &
[mag/$\sq \arcsec$] & [$\arcsec$] \\
\noalign{\smallskip}
\hline
\noalign{\smallskip}
CGF~9-6 & 145 & dE &   3:35:05.48 & -35:13:06.5 & 19.06 & 1.03 & 23.34 & 16.6 &
24.39 &  5.2 & 23.44 &  3.6 \\
CGF~9-12 & 154 & dE &  3:35:30.48 & -35:15:05.9 & 20.09 & 0.99 & 23.56 & 17.0 &
25.14 &  6.7 & 24.30 &  5.6 \\
CGF~8-1 & 160 & dE,N & 3:36:03.94 & -35:23:20.5 & 17.63 & 1.05 & 22.67 & 34.0 &
24.50 & 12.2 & 23.36 &  7.2 \\
CGF~8-3 & 162 & dE &   3:36:06.55 & -35:25:54.5 & 19.66 & 0.89 & 23.64 & 12.0 &
25.02 &  5.0 & 24.16 &  3.5 \\
CGF~6-5 &     & dE,N & 3:36:58.50 & -35:29:45.5 & 19.08 & 0.98 & 21.83 & 20.0 &
24.46 &  7.6 & 23.93 &  6.3 \\
CGF~4-1 & 202 & dE,N &     3:38:06.54 & -35:26:24.4 & 14.88 & 1.34 & 19.58 & 64.0 &
21.47 & 10.7 & 20.48 &  6.7 \\
CGF~3-1 & B1241 & dE/dS0 &3:38:16.67 & -35:30:27.9 & 16.97 & 0.93 & 21.80 & 38.0 &
23.13 &  8.7 & 22.11 &  5.3 \\
CGF~3-2 & 208 & dE,N &    3:38:18.71 & -35:31:52.1 & 17.21 & 1.12 & 20.82 & 35.6 &
23.69 &  8.5 & 22.98 &  6.6 \\
CGF~1-44 &   &   dE &     3:38:42.26 & -35:33:08.2 & 20.88 & 0.82 & 23.29 & 12.0 &
25.09 &  6.7 & 23.97 &  3.2 \\
CGF~1-4 &    &  cE/GC? &  3:38:54.05 & -35:33:33.9 & 17.87 & 1.12 & 19.14 & 11.2 &
20.03 &  1.1 &       &      \\
CGF~5-4 &  &    cE/GC? &  3:39:35.92 & -35:28:24.9 & 19.10 & 1.05 & 19.96 &  7.5 &
20.61 &  0.8 &       &      \\
CGF~10-13 & 272 & dE &  3:42:10.91 & -35:26:32.5 & 19.93 & 0.58 & 24.30 & 12.0 &
25.56 &  6.5 & 24.94 &  6.3 \\
CGF~10-11 &  &    dI? & 3:42:16.11 & -35:20:21.0 & 19.72 & 0.92 & 23.11 & 10.2 & 
23.86 &  2.9 & 22.76 &  2.1 \\
\noalign{\smallskip}
\hline
\end{tabular}
}
\end{flushleft}
\end{table*}

A growth curve analysis for each galaxy has been made with increasing elliptical
apertures using the position, ellipticity, and position angle of the
SExtractor photometry results. As local background the SExtractor value of the
interpolated sky map for each individual galaxy was taken (see Sect.~4).
In some cases overlapping stars have been removed before the analysis by
interpolating the galaxy surface brightness profile from a unaffected ring
outside the stellar profile. In other cases, the region of an overlapping star
was masked out during the fitting process of the surface brightness profile.
Two model-independent parameters have been determined for the galaxies:
the effective semi-major axis $a_{\rm eff}$ (major axis of the ellipse that
contains half of the total light) and the mean effective surface brightness
within the effective semi-major axis, $\langle \mu_{\rm eff} \rangle = V_{\rm tot}
+ 5\cdot$log$(a_{\rm eff}) + 2.5\cdot$log$(2\pi(1 - \epsilon))$, with
$\epsilon =
1 - b/a$ ($b =$ semi-minor axis, $a =$ semi-major axis).
These parameters, as well as the size of the full major axis $D_{26}$ at the
isophote of $V = 26$ mag~arcsec$^{-2}$, are given in the photometric catalog
(Appendix A).

The $\mu_{\rm eff}$-$V_{\rm tot}$ is shown in Fig.~3
(middle panel). Qualitatively, this plot is comparable to the
$\mu_{\rm peak}$-$V_{\rm tot}$ diagram in the upper panel.
The sequence of
dwarf galaxies is clearly separated from the location of background galaxies.
However, the nucleated dEs, that are hidden in the $\mu_{\rm peak}$ plot among
the background galaxies, fall in the range of the dE sequence when measuring
$\mu_{\rm eff}$. In both diagrams, there are some galaxies located below
the bulk of background galaxies, falling in the range of dwarf ellipticals.
Each individual galaxy in this region has been individually inspected.
Most of them are background spirals or galaxies that have close neighbours
which disturb the correct $\mu_{\rm eff}$ calculation. Furthermore, radial
velocity measurements have shown that nearly all galaxies at the bright
$\mu$ limit of the dE sequence are indeed background objects (see crosses
in the middle panel of Fig.~3.

The determination of the true central surface brightness $\mu_0$ is critical for
objects with small angular diameters. The centrally peaked light
distributions are blurred by seeing, which leads to a dimming of $\mu_0$.
If the apparent core radius, where the surface intensity
has decreased by a factor of 2 from its central apparent value, is smaller
than $2\sigma_\ast (\sigma_\ast = FWHM/2.354)$, the SB profile cannot be 
deconvolved from the seeing profile (Schweizer \cite{schw}, Kormendy 
\cite{korm}).
In our observations the average seeing dispersion is about $\sigma_\ast = 
0\farcs45$. 

In the following, the calculations are restricted to a subsample of galaxies,
whose apparent core radii $r_{\rm c,app}$ are larger than $0\farcs9$.
The corrections given by Kormendy (\cite{korm}) were applied to derive true 
core radii
and true central surface brightnesses $\mu_{\rm 0,cor}$.
Note that for objects with $r_{\rm c,app} = 0\farcs9$ the correction
for the true central surface brightness is of the order of 2 magnitudes,
and the true core radius is about a third of the apparent one.
At the distance of the Fornax cluster $0\farcs9$ corresponds to about 80 pc.
The Local Group dSphs, for example, have core radii between 150
and 600 pc (Caldwell et al.~\cite{cald92}, Mateo et al.~\cite{mate}). Thus, 
the apparent
surface brightnesses of dEs
in the Fornax cluster should be nearly identical with their true
central surface brightnesses, whereas compact dwarfs (like M32) and background
galaxies are severely underestimated in their measured $\mu_{\rm peak}$.
The core radius of M32, for example, is about 500 times smaller
than that of Local Group and Virgo dEs ($\simeq 1$-$2$ pc,
Kormendy \cite{korm}).
Figure 3 (lower panel) shows the $\mu_{\rm 0,cor}$-$V_{\rm tot}$
for all galaxies with corrections less than 1.5 mag.
The different symbol sizes divide
the sample in degrees of resolution of the core, as given by the ratio
$r_{\rm c,app}/\sigma_\ast$, which can directly be translated into the
correction
in magnitudes $\Delta\mu$: $\Delta\mu < 0.2$ mag corresponds to
$r_{\rm c,app}/\sigma_\ast > 5$, $\Delta\mu < 0.5$ mag to
$r_{\rm c,app}/\sigma_\ast
> 3$, and $\Delta\mu < 1.5$ mag to $r_{\rm c,app}/\sigma_\ast > 2$.

\begin{figure*}
\psfig{figure=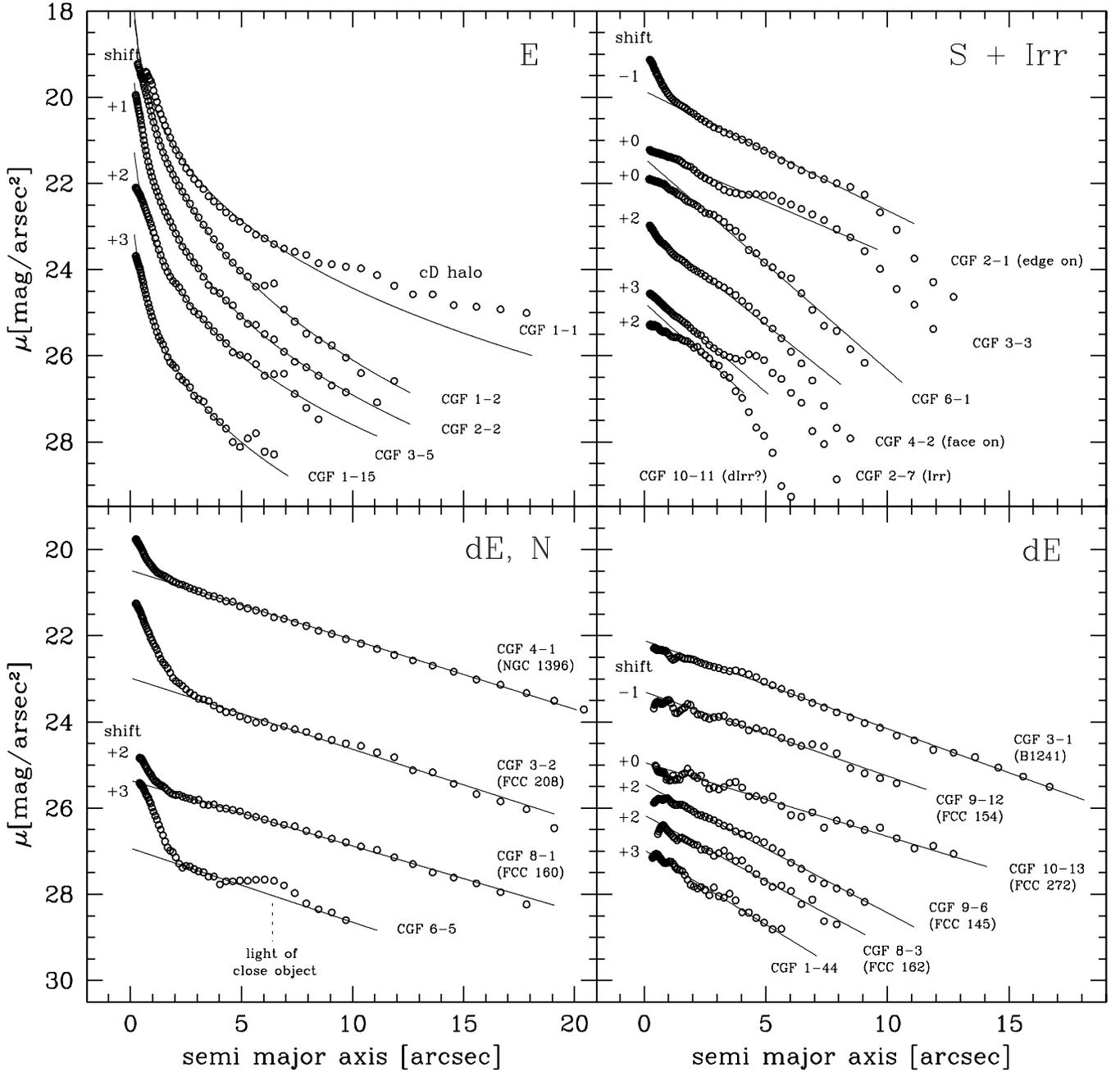,height=17.0cm,width=18.0cm
,bbllx=13mm,bblly=65mm,bburx=198mm,bbury=236mm}
\vspace{0.4cm}
\caption{Typical SB profiles of different galaxy types are
shown in the four panels. For better illustration most
profiles are shifted in $\mu$ as indicated by the numbers.
The labels are the catalog names of this paper, cross references of the
FCC catalog are given in parenthesis. In the upper left panel
ellipticals are plotted together with de Vaucouleurs profile fits. The
elliptical
CGF~1-1 has a cD halo and is the brightest galaxy of a background
galaxy cluster behind NGC~1399. The upper right panel
shows the profiles of 4 spirals and 2 irregular galaxies with exponential
fits to their disk components.
In the lower panels, all galaxies that have been classified as nucleated and
non-nucleated dEs (see Table 2) are shown
}
\end{figure*}

All non-nucleated Fornax cluster dwarfs are
clearly separated from the bulk of background galaxies and fit the sequence
of the Local Group dSphs. Their parameters are listed in Table 2. 
However, three well resolved background galaxies between
$17 < V < 19$ mag also fall in this range. Visual inspection of these objects
showed that they all are spirals seen edge-on. The surface
brightness of edge-on spirals appears to be very low due to the light 
absorption by dust in the plane of their disks.

\begin{figure}
\psfig{figure=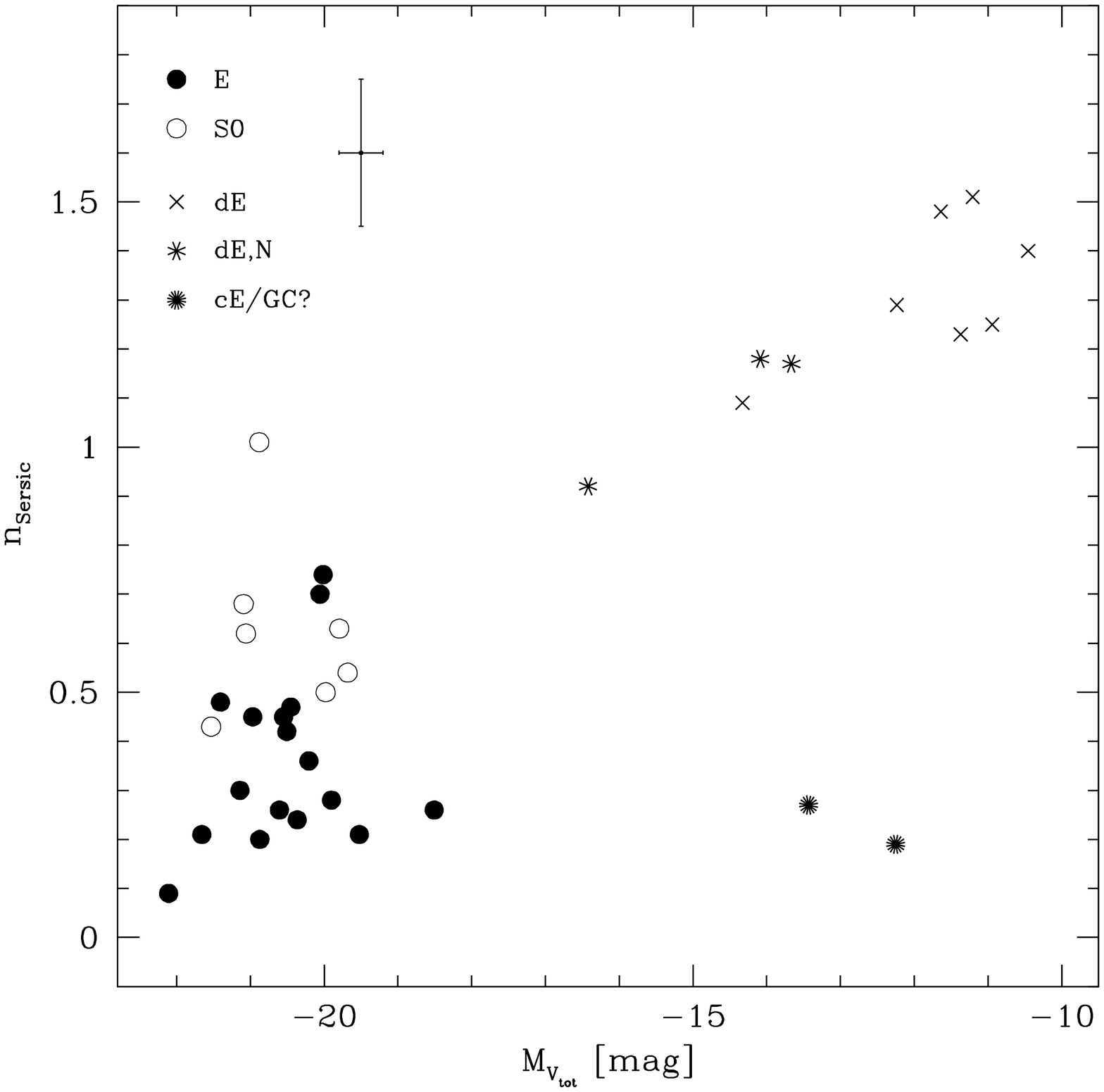,height=8.6cm,width=8.6cm
,bbllx=9mm,bblly=65mm,bburx=195mm,bbury=246mm}
\vspace{0.4cm}
\caption{The ``n'' parameter of the S\'ersic profile fits is plotted versus the 
total absolute $V$ magnitude for early-type galaxies. Filled and open circles
indicate ellipticals and S0 galaxies from our spectroscopic sample (Paper~II).
Crosses are dwarf ellipticals, asterisks nucleated ones. The two dense asterisks
in the lower right are the two compact objects found in the Fornax cluster.
A typical error for the data points is given in the upper left. For the dwarf
galaxies a n--luminosity relation is clearly visible
}
\end{figure}

Most of the galaxies that are located in the dwarf region are already listed
in the FCC (see Table 4, Appendix A, for cross references). Only 4 additional
galaxies are probably
dwarfs in the Fornax cluster due to their morphological appearance and their
photometric properties.
Their catalog names are: CGF~6-5, CGF~1-44, CGF~3-1, and CGF~10-11.

CGF~6-5 is
located close to NGC~1387 and is visible only after subtraction of the
galaxy light of NGC~1387. It looks like a nucleated dwarf elliptical judged
from its surface brightness profile (see Fig.~4 lower left panel).
CGF~1-44 is a dwarf elliptical south-east of NGC~1399. CGF~3-1 is listed in the
catalog of probable background galaxies by Ferguson (\cite{ferg}). According to
our results this galaxy more likely resembles a dwarf elliptical or dwarf
spheroidal in the Fornax cluster. CGF~10-11, located about $3\farcm5$ north
of NGC~1427, has an irregular shape and a quite blue color $(V - I) = 0.9$ mag.
According to its surface brightness it is most likely a dwarf irregular.
Nevertheless, this galaxy would be very small if it would belong to the Fornax
cluster, and therefore might be a background object.

Two objects with very high surface brightness have been identified as
Fornax members due to their radial velocity (see Paper~II). As discussed
in more detail in Paper~II, these objects might be
stripped nucleated dwarf ellipticals or very bright globular clusters.
These examples show that compact objects at the Fornax distance can hardly be
distinguished in their photometric properties from bulges of background spirals
or ellipticals. More compact Fornax objects might be hidden in our galaxy
sample. The photometric properties and profile fitting parameters of the
galaxies that has been classified as Fornax members due to their morphological
appearance, surface brightness -- magnitude relation or radial velocity are
given in Table 2.

Three different light profiles were fitted to a subsample of our galaxies
that have major axis diameters larger than $D_{26} = 7\arcsec$.
Depending on the shape of the SB profile type, an exponential law, a
$r^{1/4}$-law (de Vaucouleurs \cite{deva}) and/or a generalized exponential law
were fitted to the outer part of the profiles, which are nearly
unaffected by seeing effects. The innermost radius limit for all fits was
$1\farcs5$.
In Fig.~4 selected surface brightness profiles and their fits are shown.
The upper panels give typical examples for background ellipticals,
spirals and irregulars. The lower panels show all profiles of the Fornax
cluster nucleated and non-nucleated dEs from Table 2. The morphological types
of the galaxies were not only determined on the basis of the profiles
themselves, but also on properties like color, ellipticity, small scale 
structure (i.e. spiral arms, knots, etc.), and spectral informations
(see Paper~II). However, with decreasing angular diameter
of the galaxy the morphological classification becomes more and more uncertain.
Therefore, all type classifications with a ``?'' behind (Appendix B) should
be taken with caution; they are more a guess than a certain determination.

The ellipticals have been fitted by a de Vaucouleurs profile of the form\\
\centerline{$I(r) = I_0$exp$(-7.67(r/r_{\rm eff})^{1/4})$ or}
\centerline{$\mu(r) = \mu_0 + 8.328(r/r_{\rm eff})^{1/4}$,} 
$\mu_0$ being the central
surface brightness and $r_{\rm eff}$ the effective radius where the surface
brightness is half the central value.
In addition, we fitted the profiles of all early-type galaxies (Es, S0s, and
dEs) by the generalized exponential law (S\'ersic \cite{sers})\\
\centerline{$I(r) = I_0$exp$(-r/\alpha)^n$ or}
\centerline{$\mu(r) = \mu_0 + 1.086(r/\alpha)^n$, $n > 0$,}
which has been shown to describe the observed profiles much better (e.g.
Graham et al.~\cite{grah}). Furthermore, it is under discussion whether the
exponent ``n'' of the S\'ersic fit can be used as luminosity indicator.
Young \& Currie (\cite{youn}) found that an n--luminosity relation exists for 
early-type dwarf galaxies. Jerjen \& Binggeli (\cite{jerj}) suggested that this
relation even is continued towards ``normal'' ellipticals.
In Fig.~5 we show the ``n'' parameter plotted versus the absolute luminosity 
$M_{V_{tot}}$ for all Fornax dwarf elliptical in our sample and the 
early-type galaxies of our spectroscopic sample (see Paper~II). The absolute
luminosities were determined by adopting a distance modulus of $(m-M)_0 = 31.3$ 
mag to the Fornax dwarfs, and using the radial velocity and a Hubble constant
of $H_0 = 70$ for the other galaxies. Whereas the scatter of ``n'' for the
brighter galaxies is quite large due to the small angular diameter of the
surface brightness profiles, the n--luminosity relation for the dwarf galaxies
is clearly visible. The two compact Fornax objects do not follow the relation, 
but have ``n'' values comparable to ellipticals.

Spiral galaxies, S0s, dwarf ellipticals, and irregulars were fitted in the
outer (disk) part by an exponential law,\\
\centerline{$I(r) = I_0$exp$(-r/r_D)$ or}
\centerline{$\mu(r) = \mu_0 + 1.086(r/r_D)$,}
$r_D$ being the characteristic radius where
$I_0$ has decreased by a factor of $e^{-1}$. For some galaxies the inner
part of the profile exceeds the fitted exponential law in brightness
indicating the
presence of a bulge or nucleus. Other galaxies show a light deficiency in the
center, which might be due to seeing effects or due to dust.
All fitting parameters are summarized in a catalog, Appendix B.

\begin{figure}
\psfig{figure=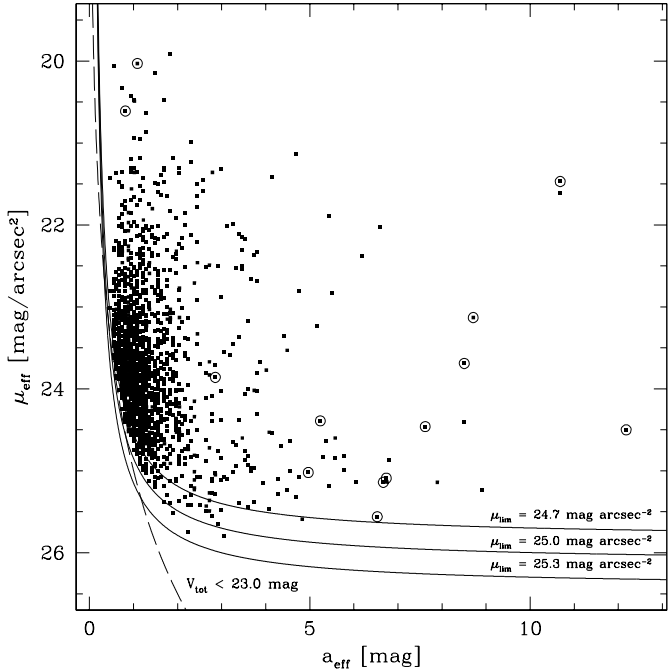,height=8.6cm,width=8.6cm
,bbllx=9mm,bblly=65mm,bburx=195mm,bbury=246mm}
\vspace{0.4cm}
\caption{The effective surface brightness of all galaxies brighter than
$V = 23$ mag is plotted versus the effective semi-major axis. Circles 
indicate Fornax members.
The dashed lines show the relation of an exponential law for the limit of the
total magnitude. The solid lines are the selection functions for our sample
for three typical detection limits (see Table 1) and a limiting radius of
$0\farcs6$. The regions below and left of these limits are inaccessible
to our survey
}
\end{figure}

\begin{figure}
\psfig{figure=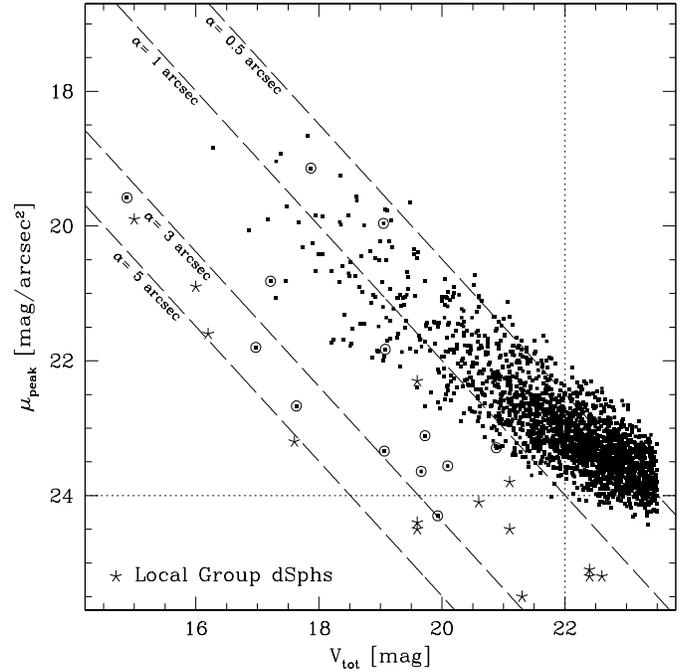,height=8.6cm,width=8.6cm
,bbllx=9mm,bblly=65mm,bburx=195mm,bbury=246mm}
\vspace{0.4cm}
\caption{The peak surface brightness of all galaxies is plotted versus the 
total V magnitude.
The dashed lines show the relation of an exponential law for different scale
lengths $\alpha$. At $V = 22$ mag the completeness starts to drop. The limit in
peak surface brightness is about $\mu_{\rm limit} = 24.0$ mag.
Circles indicate Fornax members. The asterisks are the Local
Group dwarf spheroidals shifted to the Fornax distance. None of either
category falls in the crowded part of the diagram
}
\end{figure}

\begin{figure}
\psfig{figure=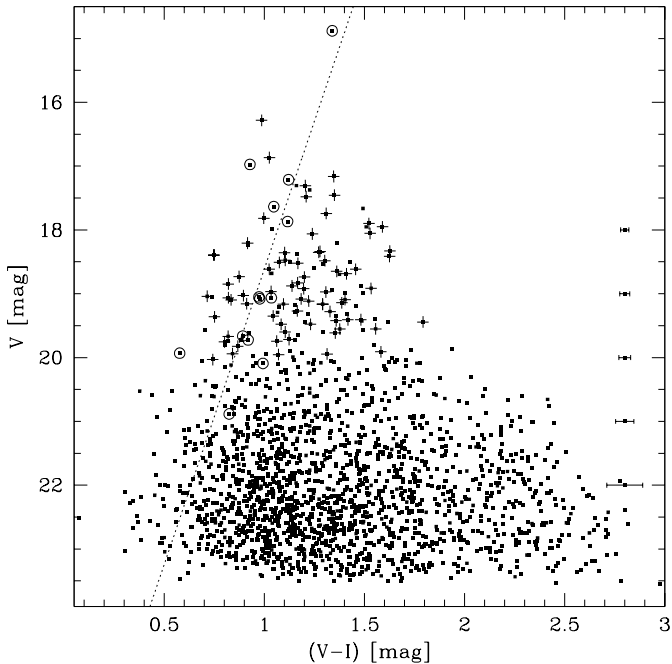,height=8.6cm,width=8.6cm
,bbllx=9mm,bblly=65mm,bburx=195mm,bbury=248mm}
\vspace{0.4cm}
\caption{The plot shows the color magnitude diagram of all galaxies in our
CCD fields. The $(V - I)$ colors were measured in apertures with
diameters of 3\arcsec. Mean errors are shown on the right.
The Fornax dwarf galaxies are encircled. 
The dashed line is a fit to 8 dEs and dE,Ns from Table~2, excluding
the bluest and reddest dwarf. A clear trend is visible in the sense that
fainter dEs are bluer, as it is also seen in other clusters (e.g. in Coma, see
Secker \& Harris \cite{seck97}). Crosses indicate definite background galaxies 
according to their radial velocities (see Paper~II)
}
\end{figure}

\section{Completeness at the Fornax distance}

The completeness of a galaxy catalog depends on the following three
parameters: limiting magnitude, limiting surface brightness, and limiting
scale length.
The surface brightness detection limits of the different fields are
given in Table 1. They vary between $24.7 < \mu_{\rm lim} < 25.3$ mag for the
central fields. The minimum number of connected pixels for a detection,
$n_{\rm min} = 5$ results in a limiting radius of about $r_{\rm lim} = 0\farcs6$.
Adopting an exponential law for the galaxy profiles, $\mu (r) = \mu_0 +
1.086(r/\alpha)$, the relation of $r_{\rm lim}$ to $\mu_{\rm lim}$ is 
$r_{\rm lim} = 0.921\cdot \alpha (\mu_{\rm lim}-\mu_0)$.
With $r_{\rm eff}=\alpha / 0.5958$ and $\mu_{\rm eff} = \mu_0 + 1.1245$,
the selection function in the $\mu_{\rm eff}$, $r_{\rm eff}$ plane is 
$\mu_{\rm eff}(r_{\rm eff}) = \mu{\rm lim}
+ 1.1245 - 1.8225 r_{\rm lim}/r_{\rm eff}$. In Fig.~6 all galaxies brighter
than $V_{\rm tot} = 23$ mag
are plotted in this plane. The selection functions for three typical
detection limits are shown. Objects that are located below and left of this 
functions are not accessible to our survey.

Figure 7 shows our sample of galaxies in a
$\mu_{\rm peak}$--$V_{\rm tot}$
diagram. Note that for most galaxies the $\mu_{\rm peak}$
is a lower limit compared to the true central surface brightness as
shown in the previous section.
Also given are the parameters of Local Group dSphs (Mateo et al.~\cite{mate})
shifted to the Fornax distance.
The limiting $\mu_{\rm peak}$ is about 24.0 mag~arcsec$^{-2}$.
For the $V$ magnitude the galaxy counts start to be incomplete
for $V_{\rm tot} > 22.0$ mag. Concerning dEs in the Fornax cluster which
follow the $\mu$-$V$ relation the completeness starts to drop at even brighter
$V_{\rm tot}$. Thus, for the Fornax dEs we are more restricted in surface
brightness than in the absolute magnitude. As shown in
Fig.~7, several Local Group dSphs would not have
been detected due to their low surface brightnesses, even if their total
luminosities would have been within our limits. The dSphs And~I and And~II,
for example, would have total magnitudes of about $V_{\rm tot} = 19.6$ mag, but
central surface brightnesses of $\mu_{\rm 0,V} = 24.5$ mag. On the contrary, the
dSph Leo~I has the same $V_{\rm tot}$, but a 2 magnitudes brighter
$\mu_{\rm 0,V}$, which is well within our sample limits.

The resolution limit is given by the seeing conditions.
All objects with FWHM larger than $1\farcs5$, or about 130 pc in Fornax
distance, appear resolved.
Thus, all Local Group dSphs
would appear clearly resolved when shifted to the Fornax distance.
The dashed lines in Fig.~7 show the limits for different scale lengths
of an exponential law in dependence of $V_{\rm tot}$ and $\mu_{\rm peak}$
surface brightness. All objects with scale lengths larger than about
$0\farcs5$ appear resolved.
%

\begin{table}[t]
\caption{Average galaxy density per square degree with errors
(considering the clustering properties of galaxies) for the different fields
in dependence on the limiting $V$ magnitude.}
\begin{flushleft}
\begin{tabular}{lrrrr}
\hline\noalign{\smallskip}
\multicolumn{2}{l}{magnitude limit:} & $V<20.0$ & $V<21.0$ & $V<22.0$\\
Field & R[$\arcmin$] & [gal.$/\sq\degr$] & [gal.$/\sq\degr$] & [gal.$/\sq\degr$]
 \\
\noalign{\smallskip}
\hline\noalign{\smallskip}
F1 (Ring1)&  2.82 & 2740$\pm$1250 & 5210$\pm$1720 & 9880$\pm$2280 \\
F2 (Ring1)&  2.82 &  960$\pm$650 & 2250$\pm$1010 & 4180$\pm$1350 \\
F3 (Ring1)&  2.82 & 1250$\pm$850 & 3330$\pm$1460 &  \\
F4 (Ring1)&  2.82 & 1040$\pm$620 & 2080$\pm$880 & 4420$\pm$1280 \\
F1 (Ring2)&  5.57 & 1170$\pm$530 & 2630$\pm$820 & 6420$\pm$1330 \\
F2 (Ring2)&  5.57 & 1830$\pm$750 & 3060$\pm$940 & 7030$\pm$1440 \\
F3 (Ring2)&  5.57 & 1490$\pm$660 & 2810$\pm$910 &  \\
F4 (Ring2)&  5.57 &  420$\pm$270 &  970$\pm$420 & 3900$\pm$920 \\
B1        & 11.54 & 1050$\pm$420 & 2200$\pm$620 & 6320$\pm$1130 \\
N1387     & 18.24 &  450$\pm$200 & 1530$\pm$410 & 6240$\pm$980 \\
N1427A    & 21.11 &  650$\pm$270 & 2100$\pm$540 & 4340$\pm$770 \\
N1379     & 28.08 &  190$\pm$120 &  830$\pm$270 & 2930$\pm$560 \\
N1374     & 41.96 &  510$\pm$220 & 1160$\pm$340 & 3210$\pm$590 \\
N1427     & 46.00 &  790$\pm$310 & 1660$\pm$460 & 4040$\pm$730 \\
B2        & 93.82 &  550$\pm$220 & 2180$\pm$530 & 4430$\pm$740 \\
B3        &611.00 &  310$\pm$ 80 & 1400$\pm$210 & 4290$\pm$420 \\
B4        &863.00 &  640$\pm$150 & 1940$\pm$310 & 5120$\pm$540 \\
\noalign{\smallskip}
\hline
\end{tabular}
\end{flushleft}
\end{table}

\section{Color distribution}

In the color magnitude diagram (Fig.~8) all galaxies brighter than $V = 23.5$
mag are shown. Most of the objects have colors of about $(V-I) = 1.2$ mag
similar to the average color of the GCSs of the
ellipticals in Fornax (Kissler-Patig et al.~\cite{kiss97}).
Assuming $E(B-V) \simeq 0$ mag towards Fornax, all galaxies
redder than $V-I = 1.8$ mag are most likely background galaxies (since no
redder stellar populations are expected from any theoretical stellar evolution
model, e.g. Worthey \cite{wort}).
The definite cluster dwarf galaxies (circles) are located at the blue side
of the color distribution. Since most of them are early-type dwarfs and show
no indications of star formation, their blue colors can only be explained
by low metallicities.
The dEs and dE,Ns follow a sequence in the sense that fainter galaxies are
bluer (see dashed line in Fig.~8). This trend was already
noticed by Caldwell \& Bothun (\cite{cald87}, who
measured $UBV$ magnitudes of 30 relatively bright Fornax cluster dwarf
ellipticals) and continues for dwarfs with low surface brightness
(Bothun et al.~\cite{both}, Cellone et al.~\cite{cell94}). Held \& Mould 
(\cite{held}) have shown
for 10 nucleated dE,Ns in Fornax that their colors are correlated with their
metallicities derived from line strengths ($-1.5 < [{\rm Fe/H}] < -0.8$).
A color -- magnitude relation for dwarf galaxies is also seen in other galaxy
clusters,
as for example in Virgo (Caldwell \cite{cald83}, Caldwell \& Bothun 
\cite{cald87}) or Coma
(Secker \cite{seck96}, Secker \& Harris \cite{seck97}).

No color dependence on the projected distance to the center of NGC~1399
is seen for the dwarfs.
Galaxies with actual or recent star formation activity which would stand
out by very blue colors cannot be identified.
However, some moderatly blue galaxies of $V-I \simeq 0.5$ are present.
Most of them are located in the background cluster behind the Fornax center
(see Paper~II) and most likely represent irregular galaxies
surrounding ellipticals in this distant cluster. In the center of the 
Fornax cluster
a prominent example of an irregular with this color is NGC~1427A (Hilker et 
al.~\cite{hilk97}).

\section{Spatial distribution of the galaxies}

\begin{figure}
\psfig{figure=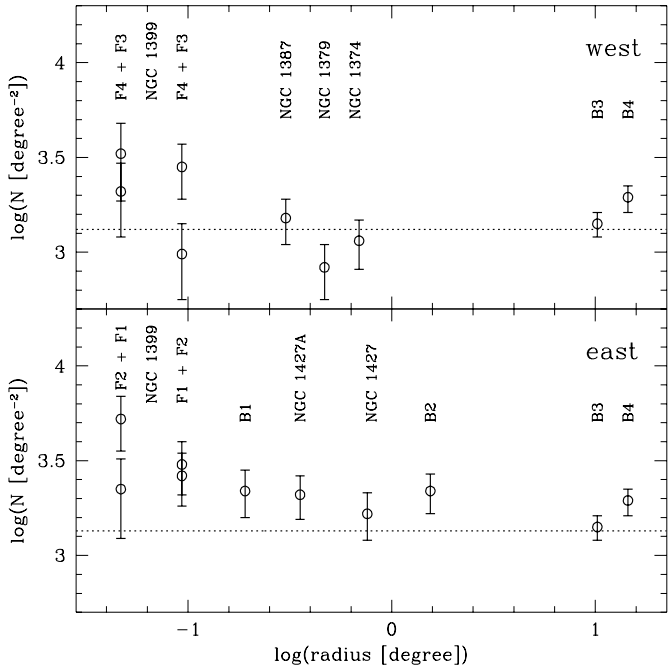,height=8.6cm,width=8.6cm
,bbllx=9mm,bblly=65mm,bburx=195mm,bbury=246mm}
\vspace{0.4cm}
\caption{The galaxy densities for galaxies brighter than $V = 21$ mag are
plotted versus the projected radial distance to the central galaxy NGC~1399.
The upper panel shows the sequence of fields in west direction, in the
lower panel the densities of fields in east direction are plotted.
In both panels
the densities of the background fields B3 and B4 are given as comparison.
An excess population of background galaxies is
seen in the central fields. The galaxy density in the eastern fields is
on the average higher than in the western fields
}
\end{figure}

We investigated the spatial distribution of all galaxies in order to
look for a possible over-abundance of dwarf galaxies in the center of the
Fornax cluster. Therefore, we calculated the galaxy densities in the different
fields.
We define in the following all CCD fields except the NGC~1399 and B1 field
as local Fornax background in contrast to the cluster center region.
The background fields B3 and B4 were used to estimate the absolute galaxy
background counts.
The galaxy counts in the central fields F1 -- F4 were divided into two rings
around the center of NGC~1399, one expanding from $0\farcm4$ to $4\farcm0$,
the other from $4\farcm0$ to $7\farcm0$.
Table 3 lists the galaxy density per square degree for the different fields
as a function of limiting $V$ magnitude. Note that for magnitudes fainter
than $V = 21$ mag the SW field of NGC~1399 (F3) has been omitted due to the
shorter exposures in this field. In all the other fields, no incompleteness
effects influence the galaxy counts for magnitudes brighter than $V = 22$
mag (see Sect.~7).

The error of the galaxy counts $N$ is a function of the angular correlation
$\omega(\theta)$ of galaxies: $\sigma^2 = N + N^2 \omega(\theta)$, $\theta$ is
the size of the field, in which the galaxies have been counted. 
Values for $\omega(\theta)$ at different limiting $V$ magnitudes
were obtained by interpolating the results in Infante \& 
Pritchet~(\cite{infa96}),
Roche et al.~(\cite{roch}) and Brainerd et al.~(\cite{brai}).
We adopted $\omega(1\degr) = 0.0115\pm0.0050$, $0.0056\pm0.0020$, and
$0.0032\pm0.0010$ for $V_{\rm limit} = 20.0$, $21.0$, and $22.0$ mag
respectively.

In Fig.~9 we plotted the galaxy densities
for galaxies brighter than $V = 21$ mag in relation to the radial
distance along the east, respectively west, sequence of our fields with 
NGC~1399 in the center. The densities
of the absolute background fields B3 and B4 are shown as comparison in 
both directions.

It is striking that the central fields, except F4 and F2 (ring~1), have 
about 2-3 times higher
density values than the other fields. There exists an excess population
of galaxies near NGC~1399. Further, the galaxy density in the eastern fields
is in the mean higher than in the western fields, whose densities are
comparable with or even lower than that of the absolute background fields.
However, nearly all density values are within the errors that result from the
density variations and clustering properties of background galaxies.
This makes it nearly impossible to distinguish a possible excess of faint
and compact dwarf
galaxies, which represent only a few percent of the galaxy counts,
from background variations.

The excess of galaxies in the central fields can be explained by
a background galaxy cluster at $z = 0.11$ just behind the center of the
Fornax cluster. In Paper~II we give a detailed analysis of radial
velocity measurements of the brightest galaxies in this region.
In Fig.~10 we show a galaxy density map of the four central Fornax fields.
We constructed this map by counting galaxies
with $16.0 < V < 21.5$ mag and $(V-I) < 1.6$ mag in bins of $150
\times 150$ pixel ($\sim 0.32$ arcmin$^2$). This sample contains 189 galaxies.
The ``density pixels'' then were
smoothed by a $3\times3$ average filter. Their values range between
0 (white) and 3.5 (black) galaxies per arcmin$^2$.
Note that the counts in the SW field (F3) and the central ``density pixel''
are not complete. In the other fields the counts are 100\% complete.
One can clearly see a banana shaped galaxy distribution east of NGC~1399.
Galaxies with observed redshifts of $z = 0.11$ (see also Paper~II) match the
distribution of high galaxy density very well. We suspect that nearly
all excess galaxies in the center belong to the background cluster. Thus,
no large excess of dwarf galaxies exists around NGC 1399.

In Fig.~11 we show the distribution of point sources around
NGC~1399. We selected all objects with classifier values larger than 0.86,
magnitudes between $20.5 < V < 22.0$, and colors in the range $0.6 < (V-I) < 
1.6$, in total 248 objects. The isodensity contours have levels between 1.1
and 6.8 objects per arcmin$^2$. The contribution of background objects was
determined by counting point sources in the local background field B2. With the 
same selection criteria 0.13 objects per arcmin$^2$ have been counted.
Thus, about 90\% of the
point sources should belong to the GCS of NGC~1399
or the background galaxy cluster. Again, the counts in all fields except F3
are 100\% complete (see also Kohle et al.~\cite{kohl}).
The peak of the distribution is located
about $1\farcm3$ east of NGC~1399. This result has to be considered
with caution,
since the counts in the central density pixels as well as in the SW field
are not complete. However, we found that this effect also (even more
pronounced) occurs when taking brighter samples of point sources, which
should not be affected by incompleteness effects.
In previous investigations of the GCS of NGC~1399, the properties of GCs in
the NE (F2) and NW (F4) field have been examined (Kissler-Patig et 
al.~\cite{kiss97}). We found
that the radial surface density profile of the GCs (centered on NGC~1399)
is shallower in the NE field than in the NW field.
Furthermore Forbes et al.~(\cite{forb}) investigated the angular distribution
of GCs within 100\arcsec ($= 5$ ``density pixels'' in a HST WFPC2 image that 
covers the NE part of the center of NGC~1399). They found a peak in the 
angular distribution in the east 
direction. Both results are consistent with our finding of an excess of 
point sources east of NGC~1399. Note that in this direction also the density
of resolved objects is very high.

\begin{figure}
\vspace{-0.1cm}
\hspace{-0.1cm}\psfig{figure=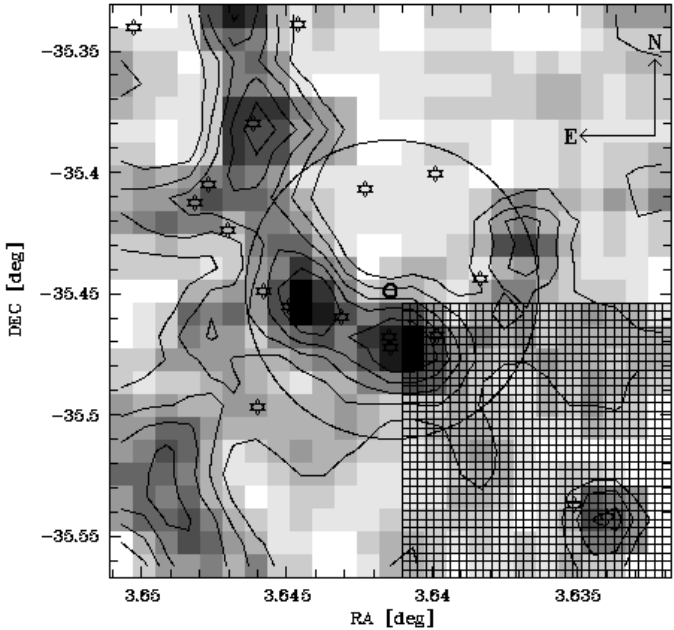,height=8.1cm,width=8.6cm
,bbllx=25mm,bblly=82mm,bburx=178mm,bbury=223mm}
\vspace{0.4cm}
\caption{The grey scale plot is a smoothed density map of galaxies brighter
than $V = 21.5$ mag in the central Fornax fields. The circle indicates
the location of NGC~1399. Asterisks are galaxies with observed redshift
of $z = 0.11$. They match the distribution of high galaxy density very well.
The contours correspond to galaxy densities of 2.3, 1.9, 1.6, 1.2, and
0.9 galaxies per arcmin$^2$.
Note that the counts in the SW field (cross-hatched region) are not complete}
\end{figure}

There are basically three possible explanations for this displacement. 
First, a significant amount of unresolved 
galaxies in the background cluster was counted together with the globular 
clusters. Since the background cluster lies east of NGC~1399 the peak
of the distribution of all point sources would then be shifted to the east.
In this case, the galaxies would have absolute magnitudes between
$-18.0 < M_V < -16.5$ mag and half light radii smaller than about 1.5 kpc
assuming a distance of 480 Mpc to the background cluster. Such properties can 
only be explained by cEs, which are believed to represent only a 
negligible fraction of the galaxy population in the local universe. 
In the Fornax cluster, for example, Drinkwater et al.~(\cite{drin97}, see also 
Drinkwater \& Gregg \cite{drin98}) investigated by
radial velocity measurements that all galaxies in their sample that are
classified as M32 type compact ellipticals in the FCC are background galaxies.
Further, in the HST counts GCs should be clearly distinguishable from galaxies
even at $z = 0.11$.
Second, the GCS is really displaced with respect to the bulge of NGC~1399. 
This would be a hint that the GCS follows another potential than the stellar 
light and may belong rather to the cluster as a whole than to NGC~1399 itself,
as supported by the velocity dispersion measurements in Grillmair et 
al.~(\cite{gril}) and Kissler-Patig et al.~(\cite{kiss98a}).
In this respect, it is worthwhile noting that the center of the gas 
distribution detected by X-ray observations is also displaced, to the 
north-east of
NGC~1399 (Ikebe et al.~\cite{ikeb}, Jones et al.~\cite{jone}).
Finally, somewhat similar to the latter point, the distribution of GCs might 
be a temporary displacement of a ``normal''
GCS centered on NGC~1399. A scenario that supports this possibility is related
to the enrichment of the central GCS by the accretion of GCs from other
galaxies. Kissler-Patig et al.~(\cite{kiss98b}) suggest that tidal tails 
of GCs from the
last passage of a Fornax galaxy might still be visible; these could mimic a 
skewed distribution of GCs around NGC~1399.
Further investigations have to show if this riddle can be solved.

\begin{figure}
\vspace{-0.1cm}
\hspace{-0.1cm}\psfig{figure=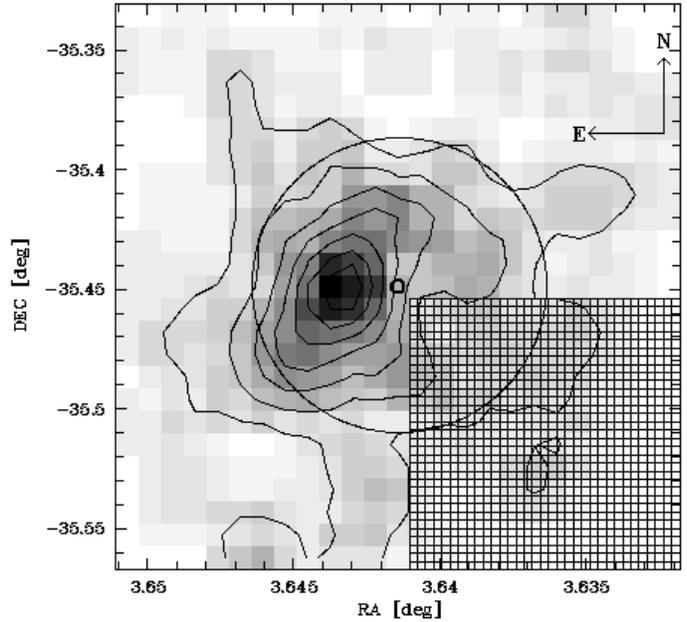,height=8.1cm,width=8.6cm
,bbllx=26mm,bblly=82mm,bburx=178mm,bbury=225mm}
\vspace{0.4cm}
\caption{The same plot as in Fig.~10 for point sources (90\% of them are
likely globular
clusters) between $20.5 < V < 22.0$. The isodensity contours correspond to
6.8, 5.8, 4.9, 3.9, 3.0, 2.1, and 1.1 objects per arcmin$^2$.
The distribution of point sources seems to be displaced by about $1\farcm3$
to the east of NGC~1399. Note that the counts in the SW field are not complete}
\end{figure}

\section{Summary}

In selected fields in the Fornax cluster, located near 
ellipticals, more than 870 galaxies were identified down to a $V$ magnitude
of 22 mag. 
Photometric properties, such as total $V$ magnitude, $(V - I)$ color, peak and
effective surface brightness, and effective radius of these galaxies are 
compiled in a catalog. 

For a bright subsample of the galaxies we determined the seeing corrected true
central surface brightnesses. Exponential and/or de Vaucouleur profiles were 
fitted
to these data and their parameters are given in a further catalog (Appendix B).
Only few galaxies could clearly be indentified as dwarf galaxies by
their location in the surface brightness magnitude diagram, where they
follow the expected sequence for dwarf ellipticals.
Most of them are already listed in the Fornax Cluster Catalog (Ferguson 
\cite{ferg}).
However, our survey limit of about 24 mag~arcsec$^{-2}$ in peak surface
brightness is too bright to detect dwarfs like the faintest Local Group dwarf 
spheroidals, if they were located at the Fornax distance.
On the other hand, among the high surface brightness objects compact dwarfs
might be hidden, as shown by two nucleus-like objects that are Fornax members
as derived from their radial velocities (see also Paper~II).

The ``n'' parameter of the S\'ersic profile fits is clearly correlated to
the absolute luminosity of the dwarf ellipticals in Fornax. The n--luminosity
relation seems to continue in the region of the ``normal'' ellipticals.

In the color magnitude diagram the dwarfs tend to follow a color -- magnitude 
relation in the sense
that fainter galaxies are bluer. The common explanation for this
relation is a decreasing metallicity with decreasing luminosity.
However, for some dwarfs the blue colors might also be a hint for an existing
young or intermediate-age stellar population, as it is seen in several
Local Group dwarf spheroidals. 
Evident signs for recent and ongoing star formation in the center of the 
cluster can only be seen in the irregular galaxy NGC~1427A.

At fixed limiting magnitude the galaxy density strongly varies from field to
field. On average the density in most fields is comparable to the one in the 
absolute background fields.
South and east of NGC~1399 we found a significant excess of galaxies as 
compared to the other fields. However, an excess of dwarf galaxies surrounding
NGC~1399 can be ruled out, since nearly all of these galaxies belong to a
background cluster at $z = 0.11$. The brightest galaxy of this background 
cluster possesses an extended cD halo and is located $1\farcm1$ south of 
NGC~1399.
The strong background galaxy fluctuations make the search for compact dwarfs
by a statistical subtraction of background objects meaningless.

The point sources in the central Fornax fields are not uniformly distributed
around NGC~1399. The peak of their density distribution is displaced about
$1\farcm3$ east of the center of the galaxy. Assuming that most of them 
are no background cluster members, but rather globular clusters, two 
explanations seem to be possible: (1) the central globular cluster system
and the bulge of NGC~1399 are disentangled from each other and
follow different potentials, or (2) tidal tales of accreted globular clusters
from passing Fornax galaxies have temporarily squewed the distribution of
globular clusters.

\begin{acknowledgements}
We thank the referee H.C.~Ferguson for his very useful comments which
improved the paper.
This research was partly supported by the DFG through the Graduiertenkolleg
`The Magellanic System and other dwarf galaxies' and through
grant Ri 418/5-1 and Ri 418/5-2. LI and HQ thanks Fondecyt Chile for support 
through `Proyeto FONDECyT 8970009' and from a 1995 Presidential Chair in 
Science.
\end{acknowledgements}
                                                 
\appendix

\section*{Appendix}

The two catalogs given in this Appendix are available in electronic form only
at the CDS via anonymous ftp to cdsar.u-strasbg.fr (130.79.128.5) or via
http://cdsweb.u-strasbg.fr/Abstract.html.
The photometric catalog contains observational data for all galaxies 
($V < 22.0$) in the 
central Fornax fields.
The second catalog contains the parameters of fits to the surface brightness
profiles of a subsample of the photometric catalog.
In both catalogs the objects are sorted in order of increasing right asccension.
In the following we describe the columns of the catalogs.

\section{The photometric catalog}

\noindent
{\bf Column 1.} Identification of the object. It is prefixed by the acronym
CGF (Catalog of Galaxies in Fornax) followed by a sequence number of the field 
and the sequence number of the galaxy in this field (ordered with decreasing
magnitude). For example, CGF~5-12 is the 12th brightest galaxy in field 5 (see
also Sect.~2.1).

\noindent
{\bf Column 2.} Right ascension for the epoch 2000 in hours, minutes and
seconds ($^h$,$^m$,$^s$).

\noindent
{\bf Column 3.} Declination (2000) in degrees, minutes and seconds 
($\degr,\arcmin,\arcsec$).

The positions of all objects were determined relative to positions in
the Guide Star Catalog. Typically 4 to 8 catalog positions are found in each
field. Coordinate transformations with 3 plate constants have been obtained.
The positional accuracy of the calculated coordinates is in all fields better
than 0.3\arcsec.

\noindent
{\bf Column 4.} Total $V$ apparent magnitude as determined by SExtractor, see
Sect.4. Values with an appended 'n' indicate that neighboring objects
are present within 2 isophotal radii.

\noindent
{\bf Column 5.} $V$ peak surface brightness in mag~arcsec$^{-2}$ as given by
SExtractor (not seeing corrected).

\noindent
{\bf Column 6.} ($V-I$) colors within an aperture of $3\arcsec$ in diameter.

\noindent
{\bf Column 7.} Ellipticity of the galaxy and the used elliptical aperture,
defined as $\epsilon = 1 - b/a$.

\noindent
{\bf Column 8.} Position angle of the major axis of the elliptical aperture.
0 degree is in east direction, positive angles towards the south, and negative
angles towards the north direction.

\noindent
{\bf Column 9.} Size of the major axis of the elliptical aperture in arcsec.
Note that the limiting isophote of the ellipse slightly varies from field
to field depending on the seeing and sky brightness.

\noindent
{\bf Column 10.} Size of the major axis at an isophotal surface brightness
of $\mu_V = 26$ mag in arcsec, $D_{26}$.

\noindent
{\bf Column 11.} Effective semi-major axis $a_{\rm eff}$ in arcsec.
Major axis containing half of the total light measured in elliptical
apertures.

\noindent
{\bf Column 12.} Effective surface brightness $\mu_{\rm eff}$, mean surface 
brightness within the effective semi-major axis.

\begin{table}
\caption{Cross references to other catalogs}
\begin{flushleft}
\begin{tabular}{llllll}
\hline\noalign{\smallskip}
 & NGC & C87 & F89 & D88\&I90 \\
\noalign{\smallskip}
\hline\noalign{\smallskip}
CGF~9-1 & & & B1016 & 231 \\
CGF~9-5 & & & FCC141 & \\
CGF~9-6 & & NG123 & FCC145 & 230 \\
CGF~9-12 & & & FCC154 & \\
CGF~8-1 & & NG16 & FCC160 & 238 \\
CGF~8-3 & & & FCC162 & \\
CGF~4-1 & 1396 & & FCC202 & \\
CGF~3-1 & & & B1241 & 95 \\
CGF~3-2 & & NG21 & FCC208 & 257 \\
CGF~3-3 & & & B1237 & \\
CGF~3-6 & & & & 258 \\
CGF~3-7 & & & B1220 & \\
CGF~1-10 & & & & 266 \\
CGF~10-2 & & & B1571 & \\
CGF~10-13 & & & FCC272 & \\
\noalign{\smallskip}
\hline
\end{tabular}
\end{flushleft}
\end{table}

\subsection{Crossreferences to other catalogs}

Some of our galaxies are also listed in previous catalogs. Table 4
gives cross references for these galaxies. The abbreviations
are as follows: C87 = Caldwell (\cite{cald87}), F89 = Ferguson (\cite{ferg},
Fornax Cluster
Catalog), D88\&I90 = Davies et al.~(\cite{davi}) and Irwin at al.~(\cite{irwi}).

\section{Catalog of profile fit parameters}

\noindent
{\bf Column 1.} Identification name as in Appendix A.

\noindent
{\bf Column 2.+3.} Right ascension and declination (epoch 2000.0) as in
Appendix A.

\noindent
{\bf Column 4.} Galaxy type as determined by morphological classification
and analysis of surface brightness profiles.

\noindent
{\bf Column 5.} Total $V$ magnitude as in Appendix A.

\noindent
{\bf Column 6.} Apparent core radius $r_{\rm c,app}$ in arcsec (= radius where 
the apparent central surface intensity has its half value).

\noindent
{\bf Column 7.} Corrected core radius $r_{\rm c,cor}$ in arcsec (using 
Kormendy's correction \cite{korm}).

\noindent
{\bf Column 8.} Corrected central surface brightness $\mu_{\rm 0,cor}$ in $V$ 
(using Kormendy's correction \cite{korm}).

\noindent
{\bf Column 9.} Exponent $n$ of Sersic profile fit.

\noindent
{\bf Column 10.} Central surface brightness $\mu_{\rm 0,Sersic}$ in $V$ of 
Sersic profile fit.

\noindent
{\bf Column 11.} Scale length $r_{\rm 0,Sersic}$ of Sersic profile fit in 
arcsec.

\noindent
{\bf Column 12.} Central surface brightness $\mu_{\rm 0,dV}$ of a de Vaucouleurs
profile fit.

\noindent
{\bf Column 13.} Effective radius $r_{\rm eff,dV}$ of the de Vaucouleurs fit in 
arcsec. Radius, where the surface brightness is two times fainter than in the 
center.

\noindent
{\bf Column 14.} Central surface brightness $\mu_{\rm 0,exp}$ of an exponential law.

\noindent
{\bf Column 15.} Scale length $\alpha$ of an exponential law in arcsec.

\noindent
{\bf Column 16.} Flag for the exponential profile fit indicating, if a bulge
``b'' is present, a light deficiency in the center ``d'', a nucleus ``n'',
or if the profile
follows an exponential law at all radii ``e''.

%


\begin{thebibliography}{}

\bibitem[1995]{bert95}
Bertin E., 1995, SExtractor 1.0 User's Guide
\bibitem[1996]{bert96}
Bertin E., Arnouts S., 1996, A\&AS 117, 393
\bibitem[1994]{bing}
Binggeli B., 1994, in: ESO/OHP Workshop on Dwarf Galaxies, eds. G. Meylan \&
P. Prugniel, ESO, Garching
\bibitem[1991]{both}
Bothun G.D., Impey C.D., Malin D.F., 1991, ApJ 376, 404
\bibitem[1995]{brai}
Brainerd T.G., Smail I., Mould J., 1995, MNRAS 275, 781
\bibitem[1991]{brid}
Bridges T.J., Hanes D.A., Harris W.E., 1991, AJ 101, 469
\bibitem[1983]{cald83}
Caldwell N., 1983, AJ 88, 804
\bibitem[1987]{cald87}
Caldwell N. L., Bothun G.D., 1987, AJ 94, 1126
\bibitem[1992]{cald92}
Caldwell N., Armandroff T., Seitzer P., Da Costa G., 1992, AJ 103, 840
\bibitem[1996]{cell96}
Cellone S.A., Forte J.C., 1996, ApJ 461, 176
\bibitem[1994]{cell94}
Cellone S.A., Forte J.C., Geisler D., 1994, ApJS 93, 397
\bibitem[1988]{davi}
Davies J.I., Phillipps S., Cawson M.G.M., Disney M.J., Kibblewhite E.J., 1988,
MNRAS 232, 239
\bibitem[1998]{dell}
Della Valle M., Kissler-Patig M., Danziger J., Storm J., 1998, MNRAS in press
\bibitem[1948]{deva}
de Vaucouleurs G., 1948, Ann. d'Astrophys. 11, 247
\bibitem[1998]{drin98}
Drinkwater M.J., Gregg M.D., 1998, MNRAS 296, L15
\bibitem[1997]{drin97}
Drinkwater M.J., Gregg M.D., Holman B.A., 1997, in ASP Conf.~Series, Vol. 116,
Proceedings of the Second Stromlo Symposium `The Nature of Elliptical Galaxies'
Eds.~M.\,Arnaboldi, G.S.\,Da Costa \& P.\,Saha
\bibitem[1989]{ferg}
Ferguson H.C., 1989, AJ 98, 367
\bibitem[1998]{forb}
Forbes D.A., Grillmair C.J., Williger G.M., Elson R.A.W., Brodie J.P., 1998,
MNRAS 293, 325
\bibitem[1996]{grah}
Graham A., Lauer T.R., Colless M., Postman M., 1996, ApJ 465, 534
\bibitem[1997]{grat}
Gratton R.G., Fusi Pecci F., Carretta E., Clementini , Corsi C.E., Lattanzi
M.G., 1997, in: "Hipparcos Venice'97 Symposium", ESA SP-402
\bibitem[1994]{gril}
Grillmair C.J., Freeman K.C., Bicknell G.V., Carter D., Couch, W.J.,
Sommer-Larsen J., Taylor K., 1994, ApJ 422, L9
\bibitem[1986]{hane}
Hanes D.A., Harris W.E., 1986, ApJ 309, 564
\bibitem[1994]{held}
Held E.V., Mould J.R., 1994, AJ 107, 1307
\bibitem[1997]{hilk97}
Hilker M., Bomans D.J., Kissler-Patig M., Infante L., 1997, A\&A 327, 562
\bibitem[1998]{hilk98ii}
Hilker M., Infante L., Vieira G., Kissler-Patig M., Richtler, T., 1998, A\&AS, 
(Paper~II)
\bibitem[1998]{hilk98iii}
Hilker M., Richtler, T., Infante L., Kissler-Patig M., 1998, A\&A, in 
preparation (Paper~III)
\bibitem[1996]{ikeb}
Ikebe Y., Ezawa H., Fukazawa Y. et al., 1996, Nature 379, 427
\bibitem[1987]{infa87}
Infante L., 1987, A\&A 183, 177
\bibitem[1996]{infa96}
Infante L., Pritchet C.J., 1996, ApJ, 439, 565
\bibitem[1990]{irwi}
Irwin M.J., Davies J.I., Disney M.J., Phillipps S., 1990, MNRAS 245, 289
\bibitem[1997]{jerj}
Jerjen H., Binggeli B., 1997, in ASP Conf.~Series, Vol. 116, Proceedings of
the Second Stromlo Symposium `The Nature of Elliptical Galaxies' 
Eds.~M.\,Arnaboldi, G.S.\,Da Costa \& P.\,Saha
\bibitem[1997]{jone}
Jones C., Stern C., Forman W., Breen J., David L., Tucker W., Franx M., 1997,
ApJ 482, 143
\bibitem[1988]{kill}
Killeen N.E.B., Bicknell G.V., 1988, ApJ 325,165
\bibitem[1997]{kiss97}
Kissler-Patig M., Kohle S., Hilker M., Richtler T., Infante L., Qintana H., 
1997, A\&A 319, 470
\bibitem[1998a]{kiss98a}
Kissler-Patig M., Brodie P.B., Schroder L.L., Forbes D.A., Grillmair C.J.,
Huchra J.A., 1998a, AJ 115, 105
\bibitem[1998b]{kiss98b}
Kissler-Patig M., et al., 1998b, AJ, in preparation
\bibitem[1996]{kohl}
Kohle S., Kissler-Patig M., Hilker M., Richtler T., Infante L., Quintana H.,
1996, A\&A 309, L39
\bibitem[1985]{korm}
Kormendy J., 1985, ApJ 295, 73
\bibitem[1980]{kron}
Kron R.G., 1980, ApJS 43, 305
\bibitem[1992]{land}
Landolt A.U., 1992, AJ 104, 340
\bibitem[1997]{lope}
L\'opez-Cruz O., Yee H.K.C., Brown J.P., Jones C., Forman W., 1997, ApJ 475, L97
\bibitem[1993]{mate}
Mateo M., Olszewski E.W., Pryor C., Welch D.L., Fischer P., 1993, AJ 105, 510
\bibitem[1993]{prug}
Prugniel P., Bica E., Klotz A., Alloin D., 1993, A\&AS 98, 229
\bibitem[1993]{roch}
Roche N., Shanks N., Metcalfe N., Fong R., 1993, MNRAS 262, 360
\bibitem[1984]{sand}
Sandage A., Binggeli B., 1984, AJ 89, 919
\bibitem[1986]{scho}
Schombert J.M., 1986, ApJS 60, 603
\bibitem[1981]{schw}
Schweizer F., 1981, AJ 86, 662
\bibitem[1996]{seck96}
Secker J., 1996, ApJ 469, L81
\bibitem[1997]{seck97}
Secker J., Harris W.E., 1997, ApJ 469, 623
\bibitem[1968]{sers}
S\'ersic J.-L., 1968, Atlas de galaxias australes, Observatorio Astronomica,
Cordoba
\bibitem[1991]{wagn}
Wagner S.J., Richtler T., Hopp U., 1991, A\&A 241, 399
\bibitem[1994]{wort}
Worthey G., 1994, ApJS 95, 107
\bibitem[1994]{youn}
Young C.K., Currie M.J., 1994, MNRAS 268, L11

\end{thebibliography}
\enddocument